\documentclass[]{iopjournal}

\usepackage{graphicx}% Include figure files
\usepackage{bm}% bold math
\usepackage{xcolor}

\usepackage{ragged2e}   % fornece \justifying
\usepackage[normalem]{ulem}
\usepackage{amsmath,amssymb} 
\usepackage{float}

\begin{document}
\justifying

\title{Quantum rotation sensing from superintegrability}

\author{Leandro Hayato Ymai$^1$\orcid{0000-0001-8427-6781},
Karin Wittmann Wilsmann$^2$\orcid{0000-0002-9203-910X},
Joel Bacellar Neves$^2$\orcid{0000-0003-1399-246X},
Arlei Prestes Tonel$^1$\orcid{0000-0002-6117-3883},
Jon Links$^3$\orcid{0000-0003-1049-0616} and 
Angela Foerster$^2$\orcid{0000-0001-6486-4164}}

\affil{$^1$Campus Bagé, Universidade Federal do Pampa, Bagé, RS, Brazil},
\affil{$^2$Instituto de F\'{i}sica, Universidade Federal do Rio Grande do Sul, Porto Alegre, RS, Brazil},
\affil{$^3$School of Mathematics and Physics, The University of Queensland, Brisbane, Australia}

\email{leandroymai@unipampa.edu.br and  karin.wittmann@ufrgs.br}

\keywords{superintegrability, quantum sensing, Bethe Ansatz}

\begin{abstract}
\justifying{
Detection based on quantum principles such as entanglement has the capacity to achieve finessed levels of sensitivity, bringing transformative impacts to applications. In this study, we propose a rotation sensing protocol using an extended Bose-Hubbard model in a four-site configuration. The design, based on a simple population imbalance measurement to quantify rotation, profits from the property of superintegrability. The implementation of the sensing protocol achieves rotation-detection sensitivity scaling as $N^{-3/2}$, where $N$ is the number of particles. Remarkably, the sensitivity lies close to the quantum Cramér–Rao bound. Our results illustrate superintegrability opportunities for advancing the field of quantum sensing.}
\end{abstract}

\section{Introduction}\label{intro} 
Metrology is a cornerstone of science and engineering, striving to obtain the highest precision in the estimation of a given parameter. 
In this work, we investigate a means to accurately estimate the angular velocity of the reference frame for a system of interacting bosons. 
We propose a protocol that exploits the property of superintegrability, i.e., where the number of independent
conserved quantities of the system is greater than the number of degrees of
freedom. This elegant feature, which has been the subject of a rich mathematical study recently~\cite{Murray2020,Pozsgay2022,Pozsgay2024,Bennett2024,YmaiNewton2025}  is rarely encountered in quantum many-body physics. However, it is a potentially powerful framework to work in because superintegrable systems display robustness against perturbations that break the superintegrability, but leave integrability in place. From this perspective, superintegrability may guide the design of sensing protocols that can be analyzed using the many analytic tools that integrability provides.   

Superintegrability was recently identified in a class of boson models that admit Bethe Ansatz solutions \cite{Bennett2024}. These models are specialisations of {\it Extended Bose-Hubbard models} that have been widely implemented for studies involving ultra-cold dipolar atoms \cite{Baranov2008,Lahaye2009,Baier_2016,Chanda2025}.  The simplest superintegrable example from \cite{Bennett2024} 
is a four-well (or equivalently, four-site) configuration. 
The four wells comprise an apex well connected by tunneling to three coplanar wells forming a triangular base (see Fig.~1).

\begin{figure}[t]
\vspace{0.2cm}
\center
          \includegraphics[width=0.6\linewidth]{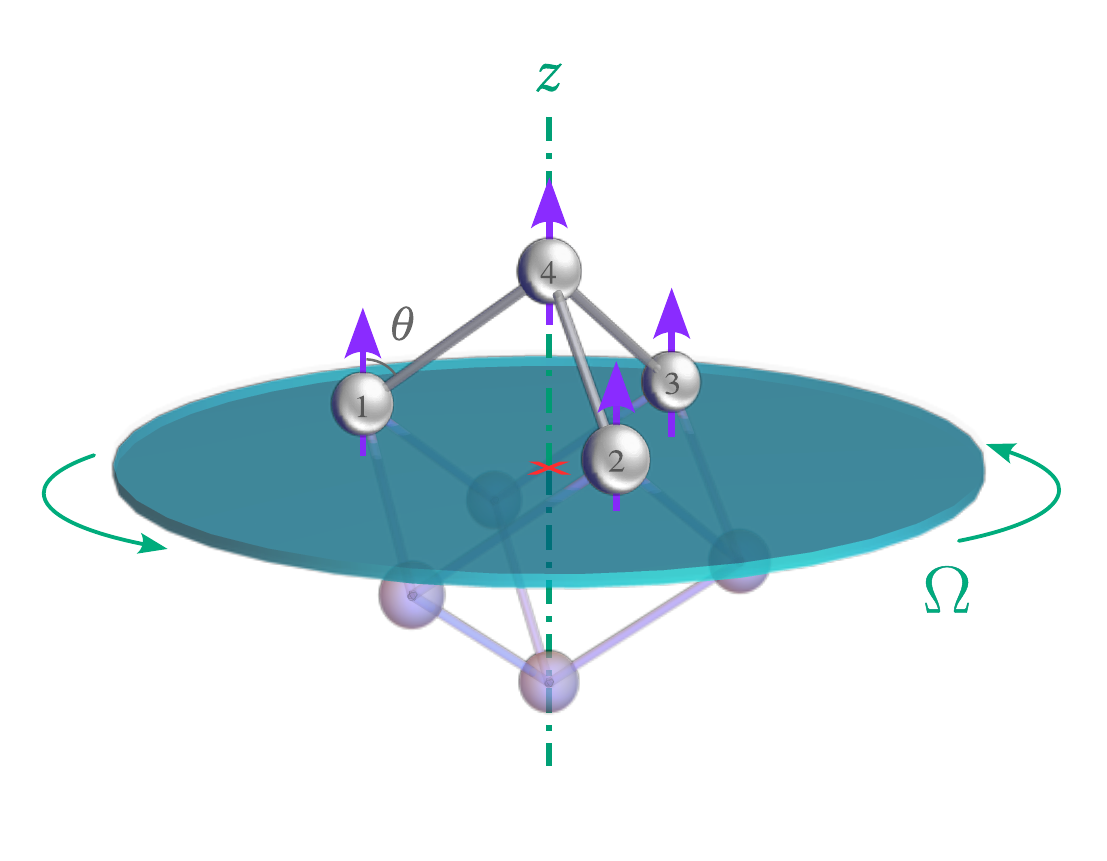}
\caption{
Schematic figure. The four-well system is embedded in a cubic lattice, and confined by an external field (not shown).
The purple arrows represent the dipoles aligned along the $z$ axis (dash-dotted line). The system is designed to sense rotations with angular velocity $\Omega$.
}
\label{fig1}
\end{figure}

For an initial state where all atoms occupy one of the outer sites, site population expectation values for the other outer sites are symmetric in a non-rotating frame. However, in a rotating frame the expectation values are skewed, and vary periodically over time. This is analogous to the phase shift induced by frame rotation observed in the Sagnac effect~\cite{Pascoli_2017}. The reference frame rotation can be directly estimated by measuring the population imbalance between  sites. Due to the underlying integrability, analytic formulae are obtained to guide the design of a rotation sensing protocol.

In Section 2 we define the model and identify the symmetries.
In Section 3 we analyse the dynamics of the system, and entanglement
 generation through the Von Neumann entropy.
Based on these results,  Section 4 presents the design of a rotation sensing protocol.  
Section 5 is devoted to the concluding remarks.
The appendices follow, containing important technical details.

\section{\label{theretical} Model and symmetries}
We consider a system of dipolar bosons confined in a four-well potential, with a central site connected to three other outer sites. This configuration can be implemented using a cubic lattice, as shown in Fig. \ref{fig1}. The system is described by the Hamiltonian of the Extended Bose-Hubbard model~\cite{Baranov2008,Lahaye2009}
\begin{eqnarray}
H_0 = \frac{U_0}{2}\sum_{i=1}^4N_i(N_i-1)+\sum_{i,j=1}^4\frac{U_{ij}}{2}N_iN_j 
%\nonumber \\ &&\qquad 
+
%\sum_{i=1}^4\nu_iN_i
J\sum_{i=1}^3(a_i^\dagger a_4 + a_4^\dagger a_i),
\end{eqnarray}
where $a_i^\dagger$, $a_i$,  and $N_i$ are the boson creation, annihilation, and particle number operators at the site $i$, respectively, with the total number of particles $N=N_1+N_2+N_3+N_4$ conserved. 
The parameter  
$J$ is the hopping rate between nearest-neighbour sites.
The on-site interaction energy $U_0$ results from the contact interaction, characterized by the scattering length, and dipole-dipole interaction (DDI) between particles at the same site, while $U_{ij}\propto (1-3\cos^2\theta)$ is the energy interaction resulting from DDI between atoms at different sites $i$ and $j$, where $\theta$ is the angle between dipole polarization and the relative position vector between site $i$ and site $j$.  The DDI obeys an inverse cubic law, and the on-site interaction energy is strongly influenced by the geometry of the potential well and the direction of dipole polarization. Here, we assume a spherical potential so that the energy contribution resulting from the on-site DDI is cancelled out~\cite{Lahaye2009} and the on-site interaction energy $U_0$ can be controlled by the scattering length via Feshbach resonance. We consider all dipoles polarized along the $z$-axis by a magnetic field, and due to the arrangement of the wells, the inter-site interaction energies satisfy $U_{14}=U_{24}=U_{34}=0$, since $\cos\theta=1/\sqrt{3}$, and $U_{12}=U_{23}=U_{13}$ by symmetry. Our objective is to analyse this system in a non-inertial frame, rotating around the $z$-axis with constant angular velocity $\Omega$. The rotational kinetic energy $H_{\rm RF} = -{\bf \Omega}\cdot {\bf L}$, depending on ${\bf \Omega}$ and angular moment ${\bf L}$, also contributes to the Hamiltonian, resulting in $H= H_0+H_{\rm RF}$~\cite{Fetter2009}. 

The superintegrable regime~\cite{Bennett2024} at $\Omega = 0$, characterized by the existence of more conservative operators than degrees of freedom of the system, is achieved by balancing the interaction energies, $U_{12} = U_0$. Under these conditions, the rotating system is described by a Hamiltonian in reduced form (see Appendix \ref{appA0} for details)     
\begin{eqnarray}
H=U(N_1+N_2+N_3-N_4)^2 -Ja_4^\dagger (a_1+a_2+a_3)  
%\nonumber\\&&
-J(a_1^\dagger+a_2^\dagger+a_3^\dagger)a_4-\zeta\mathcal{J}
,\label{H1}
\end{eqnarray}
where we define the effective interaction energy $U=U_0/4$. 
The rotation is described by the parameter $\zeta\propto \Omega$, and the current operator  given by \cite{YmaiNewton2025}
\begin{eqnarray}
\mathcal{J} = \frac{i}{\sqrt{3}}[(a_2^\dagger a_1+a_3^\dagger a_2+a_1^\dagger a_3)-(a_1^\dagger a_2+a_3^\dagger a_1+a_2^\dagger a_3)].\nonumber
\end{eqnarray} 

Defining the collective modes $b_k = (a_1+\nu^{k-1} a_2+\nu^{1-k} a_3)/\sqrt{3}$, where $\nu=e^{2\pi i/3}$, we find that the operators $b_j^\dagger b_k$, with $j,k=\{2,3\}$, commute with the Hamiltonian (\ref{H1}) for the non-rotating case, establishing  superintegrability \cite{Bennett2024,YmaiNewton2025}.
Remarkably, rotation breaks superintegrability but preserves integrability of the system. 
The restricted set of operators $Q_{k} =b_k^\dagger b_k$, $k=2,3$, given explicitly by
\begin{align}
Q_2 &= \frac{1}{3}\left(N_1+N_2+N_3+\nu a_1^\dagger a_2+ \nu^{-1}a_2^\dagger a_1\right.\nonumber\\
&\qquad \left. \quad+\nu^{-1}a_1^\dagger a_3+ \nu a_3^\dagger a_1+\nu a_2^\dagger a_3+ \nu^{-1}a_3^\dagger a_2\right),\nonumber\\
Q_3 &=\frac{1}{3}\left(N_1+N_2+N_3+\nu^{-1} a_1^\dagger a_2+ \nu a_2^\dagger a_1\right.\nonumber\\
&\qquad \left.\quad +\nu a_1^\dagger a_3+ \nu^{-1} a_3^\dagger a_1+\nu^{-1} a_2^\dagger a_3+ \nu _3^\dagger a_2\right),\nonumber
\end{align}
commute with the Hamiltonian of the rotating system.  
The set of four independent commuting operators, $\{H, N, Q_2, Q_3\}$, for this system of four modes, ensures  integrability. In particular, the current operator $\mathcal{J} =Q_3-Q_2$ is also conserved. Associated with the integrability, is the existence of a Bethe Ansatz solution \cite{baxter2007exactly,Murray2013,Batchelor2016,Ymai_2017,Guan2022,Retore_2022} for the model, which is derived in Appendix \ref{appD0}.

%--------------------------------------%

\section{\label{dynmics} Dynamics and entanglement}
\begin{figure}[ht]
\center
\includegraphics[width=0.6\linewidth]{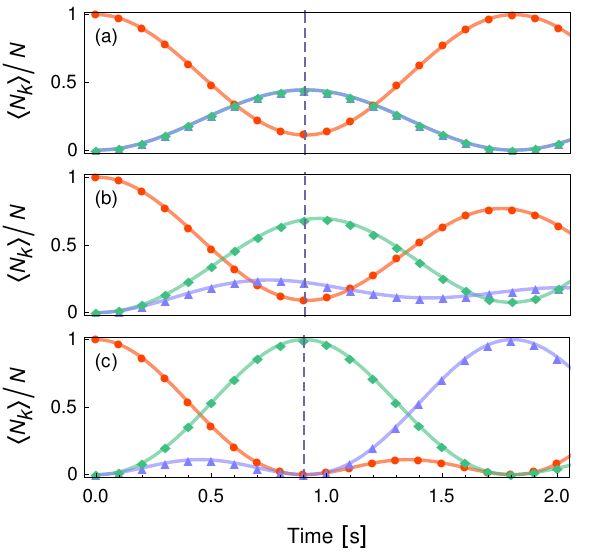}
\caption{Fractional population dynamics. Comparison between analytic expression (solid) and numerical simulation (markers) of fractional populations $\langle N_1\rangle/N$ (red),  $\langle N_2\rangle/N$ (blue) and  $\langle N_3\rangle/N$ (green), for $\zeta/h = 0$ (top), $\zeta/h = 0.3\, \zeta_{\rm max}/h$ (middle) and  $\zeta/h =\zeta_{\rm max}/h$ (bottom), where $\zeta_{\rm max} = \xi/3$. Vertical dashed line represents $t=\tau$. In all cases we use $U/h = 6.01$ Hz and $J/h= 8.16$ Hz and $N=16$.} 
\label{figQD}
\end{figure}
In this section, we analyze the influence of rotation on the dynamics of site populations of the system. This is studied through the expectation values $\langle N_k\rangle \equiv \langle \Psi(t)|N_k|\Psi(t)\rangle$, where the time evolution of the system is given by
\begin{eqnarray}
|\Psi(t)\rangle =\exp(-it H)|\Psi(0)\rangle,\nonumber
\end{eqnarray}
(we consider $\hbar=1$).
Throughout the text, we assume all dipolar bosons are initially in well 1, $|\Psi(0)\rangle = |N,0,0,0\rangle$.

The choice of coupling parameters leads to  resonant behavior of the atomic populations, where atoms tunnel between the outer wells via a virtual particle that passes to the central well due to second-order processes. This occurs when the condition 
\begin{equation}  2U(N-1) \gg J
\label{rtc}
\end{equation}
is satisfied, leading to harmonic dynamics with period $T=2\tau$ where 
\begin{eqnarray} 
\tau= \frac{\pi}{\xi},\qquad 
\xi =\frac{3J^2}{4U(N-1) }. 
\label{xi}
\end{eqnarray}
In this resonant regime, an effective Hamiltonian can be written in terms of the conserved operators of the system, given by (see details in Appendix \ref{appD0})

\begin{eqnarray}
H_{\rm eff}=-( \xi-\zeta) Q_2-( \xi+\zeta) Q_3.\nonumber
\end{eqnarray}
Using the effective Hamiltonian in the time evolution operator, it is observed that the state of the system is  a coherent state
\begin{eqnarray}
|\Psi(t)\rangle=\frac{[\sigma_1(t) a_1^\dagger +\sigma_2(t) a_2^\dagger+\sigma_3(t) a_3^\dagger]^N|0\rangle}{3^N\sqrt{N!}},\label{psi}
\label{coherent_state}
\end{eqnarray}
where $\sigma_k(t) = 1+2e^{-it\xi}\cos\theta_{k}(t)$ and $\theta_{k}(t)=t \zeta+2\pi(k-1)/{3}$. This leads to an analytic expression for expectation values of populations, given by (for $k=1,2,3)$
\begin{eqnarray}
\langle N_{k}\rangle = \frac{N}{9} \left( 1+4\cos\theta_{k}(t) \left[\cos(\xi t)+\cos\theta_{k}(t) \right] \right ) .
\label{number}
\end{eqnarray}

The parameter values used in this study were chosen to satisfy the integrability constraints and resonant conditions of the model.
Fig. \ref{figQD} shows the comparison between the analytic results and the numerical simulation of fractional populations of sites 1, 2, and 3 for different values of $\zeta$.  
Fig. \ref{figQD}(a) shows the configuration without rotation, where the initial population in well 1 divides in the same proportion between wells 2 and 3 after a time interval $t =\tau$. On the other hand, Figs. \ref{figQD}(b) and  \ref{figQD}(c) show that as the rotation speed increases, the proportion between the populations of wells 2 and 3 at $t=\tau$ becomes progressively more unbalanced, reaching a maximum difference when $\zeta = \zeta_{\rm max}\equiv \xi/3$, as observed in Fig. \ref{figQD}(c).
At $t = \tau$, the formula for the imbalance population obtained from (\ref{number}) reads
\begin{eqnarray}
\left.\langle N_2-N_3\rangle \right|_{t=\tau} = \frac{2N}{3\sqrt{3}}\left[\sin(2\pi\zeta/\xi)+2\sin(\zeta \pi/\xi)\right].\label{imb}
\end{eqnarray}

In the absence of rotation, the state of well 3 is highly entangled with the rest of the system at $t=\tau$. 
This is verified through the Von Neumann entropy $S=-{\rm Tr}(\rho\log\rho)$, where the reduced matrix $\rho = {\rm Tr_{124}}|\Psi(t)\rangle\langle \Psi(t)|$ is obtained by tracing over the subsystem of wells 1, 2 and 4 (see details in Appendix \ref{appE0}). Fig. \ref{fig: S} shows the variation of the entropy under rotation as $\zeta$ increases from zero to $\zeta_{\rm max}$. We observe that the entropy decreases monotonically with $\zeta$ from its maximum to zero. This is due to the passage from the output coherent state $(-a_1^\dagger+2a_2^\dagger+2a_3^\dagger )^N|0,0,0,0\rangle$ at $\zeta=0$ to the output Fock state $  |0, N,0,0\rangle$ at $\zeta=\zeta_{\rm max}$, as given by (\ref{coherent_state}).
\begin{figure}[!ht]
\center
\includegraphics[width=0.6\linewidth]{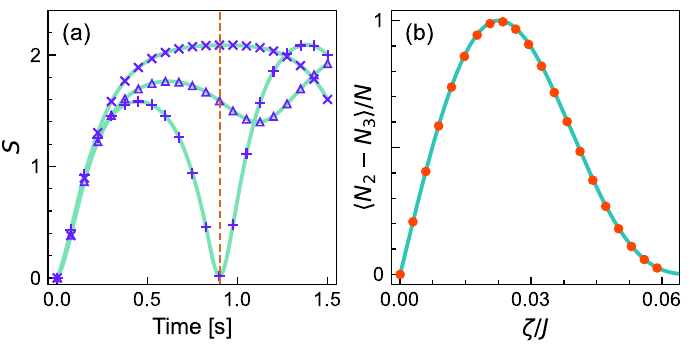}
\caption{(a) Entanglement dynamics of well 3. Comparison between analytic expression (solid) and numerical simulation (markers) of the Von Neumann entropy for 
$\zeta = 0$ ($\times$),  $\zeta = \zeta_{\rm max}/2$ ($\triangle$) , and  $\zeta = \zeta_{\rm max}$ (+).
Vertical dashed line represents $t=\tau$. (b) Imbalance population as function of $\zeta/J$. Vertical dashed line represents $\zeta/J= \zeta_{\rm max}/J$. In all cases we consider $U/h = 6.01$ Hz, $J/h= 8.16$ Hz and $N=16$.} 
\label{fig: S}
\end{figure}

%---------------------------------------%

\begin{figure*}[t]
\centering
\begin{minipage}[c]{0.7\textwidth}
    \centering
        \includegraphics[width=\linewidth]{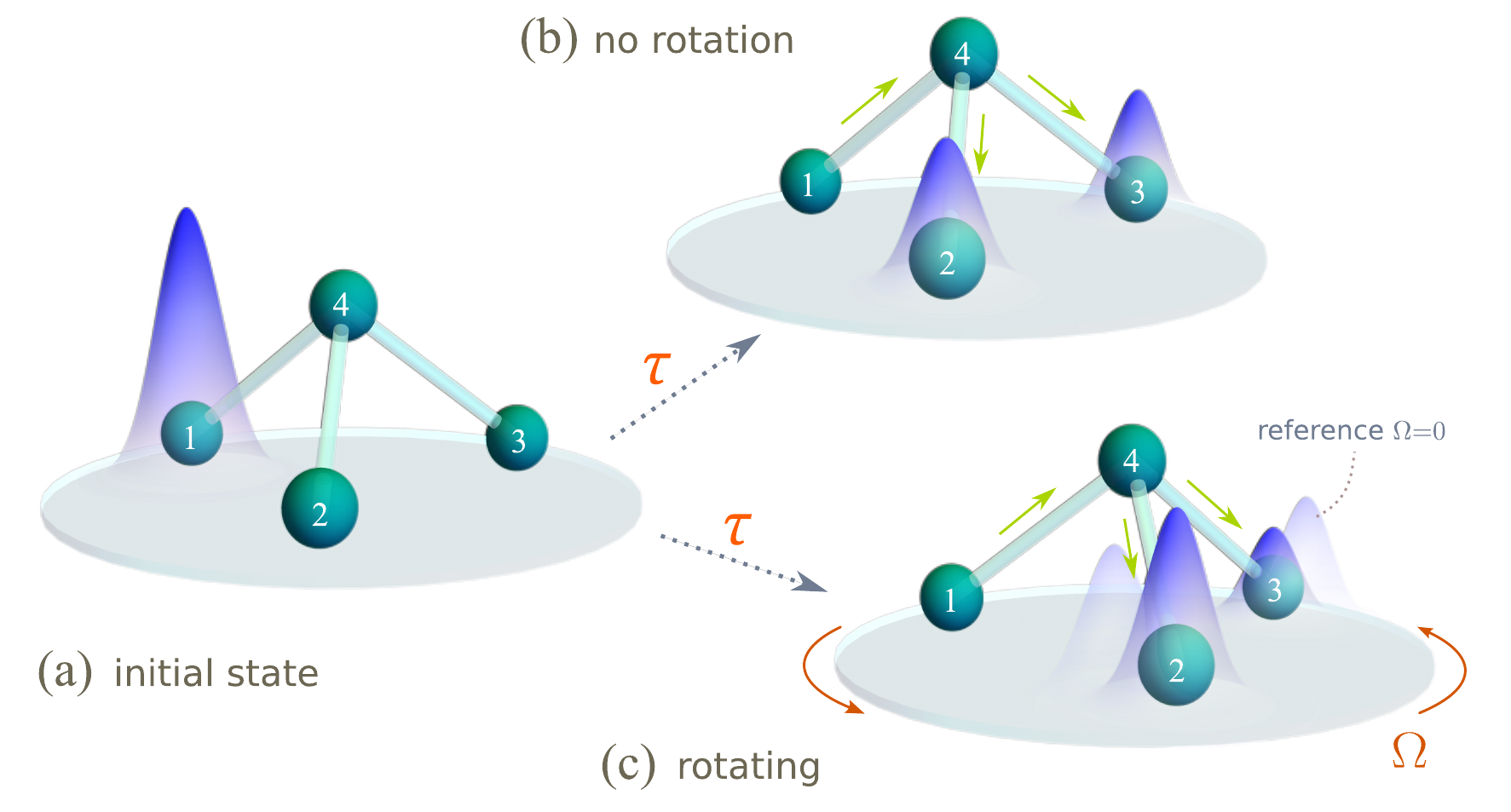}
\end{minipage}\hfill
\begin{minipage}[c]{0.27\textwidth}
    \footnotesize
\caption{\justifying Rotation sensing protocol. (a) The population is initially localized at site 1 and evolves for a time \(\tau\). (b) In the absence of rotation, it is equally distributed between sites 2 and 3. (c) In the presence of rotation, an imbalance develops between sites 2 and 3. The population imbalance provides a measure of the angular velocity. In both cases, tunneling occurs via virtual processes mediated by site 4.}
    \label{fig2}
\end{minipage}
\end{figure*}

\section{\label{sensor} Design of a rotation sensing protocol}
\noindent In order to design a rotation sensing protocol, we need to first identify an appropriate 
{\it interrogation time} to implement a measurement. Due to the harmonic nature of dynamics in the resonant regime, along with the observations of the entanglement dynamics above, it is optimal to choose the interrogation time to be the half period $\tau$ given by (\ref{xi}). 
From the analysis in the previous section, we find that the expectation values of populations at sites 2 and 3 when $t = \tau$, $\langle N_2\rangle$ and   $\langle N_3\rangle$, are affected by the rotation of the system. The imbalance population between sites 2 and 3 quantifies the magnitude of rotation,
due to the one-to-one correspondence between the intensity of the imbalance population and the angular velocity in the interval
$0 \leq \zeta \leq \zeta_{\rm max}$. Fig.~\ref{figQD} depicts  the relationship. This feature underpins the utility of the system for rotation sensing, operating over this interval. See Fig.~\ref{fig2} for an illustration.

%%%%%%%%%%%%%%%%%%%%%%%%%%%%%%%%%%%%%%%%%%%%%%%%

It is convenient to analyze the rotation sensitivity in terms of a dimensionless parameter related to angular velocity. To this end, we define
\begin{eqnarray}
\alpha = \zeta/J.\nonumber
\end{eqnarray}
The rotation sensitivity is evaluated using the error formula (see Appendix \ref{appF0})
\begin{align}
\Delta \alpha &= \frac{\Delta\langle N_2-N_3\rangle }{\left|\partial_{\alpha}\langle N_2-N_3\rangle\right|} \nonumber\\
&= \frac{\xi\, f(\zeta) }{2  \pi  \sqrt{2N}J},\label{error}
\end{align}
where $\Delta \langle N_2-N_3\rangle$ is the standard deviation, and $f(\zeta) =\sqrt{2+\cos \left({\pi  \zeta }/{\xi }\right)} \,\sec \left({\pi\zeta }/({2 \xi })\right)$ is a smooth, convex curve with little variation ($\sqrt{3}\leq f(\zeta)\leq \sqrt{3+1/3}$) over the interval $0\leq \zeta\leq \zeta_{\rm max}$.

Substituting Eq. (\ref{xi}) into Eq. (\ref{error}) gives 
\begin{equation}
\Delta \alpha=     \frac{3Jf(\zeta)}{8\pi U\sqrt{2N}(N-1)}
\nonumber
\end{equation}
exhibiting $\sim N^{-3/2}$ scaling.
This scaling arises from the $N$-dependence of the interrogation time.
Finally, a quantum Fisher information analysis yields the Cramér–Rao bound (\ref{crb}) (see Appendix \ref{appCR}):
\begin{eqnarray}
(\Delta \alpha)_{\rm CRB} = \frac{3\sqrt{3}J}{8\pi U\sqrt{2N}(N-1)}.\nonumber
\end{eqnarray}
An important feature of our results is that the sensitivity remains close to the Cramér-Rao bound throughout the interval  $0 \leq \zeta \leq \zeta_{\rm max}$.

%--------------------------------------%

\section{\label{conclusion} Conclusion}

In this study we demonstrated that rotation of a certain superintegrable system breaks the superintegrability whilst maintaining integrability. This system described the interaction of ultra-cold dipolar atoms in a four-well configuration embedded in a cubic lattice. 
The property of integrability  allows for the derivation of analytic formulae for the quantum dynamics and entanglement through the Bethe Ansatz solution. 

These results
were exploited to design a rotation sensing protocol, exhibiting several attractive features: the initial state in the protocol is a Fock state, the measurement at interrogation time is a simple population imbalance, and a quantum Fisher information analysis shows that the protocol operates with sensitivity close to the maximum quantified by the Cramér–Rao bound.    These results highlight the potential of the proposed quantum protocol for high-precision rotation estimation. 

%%%%%%%%%%%%%%%%%%%%%%%%%%%%%%%%%%%%%%%%%%%%%

\ack{AF and JL were supported by the Australian Research Council through Discovery Project
DP200101339, ``Quantum control designed from broken integrability''. JL acknowledges the traditional owners of the Turrbal and Jagera country on which The University of Queensland (St Lucia Campus) operates.
We would like to express our sincere gratitude to Ricardo R. B. Correia and Daniel S. Grün for their thorough review and valuable contributions.
Part of this work was conducted while AF and JL visited the Innovation Academy for Precision Measurement Science and Technology,
Chinese Academy of Sciences, Wuhan. We thank Xiwen Guan for his kind hospitality during this visit and helpful feedback on our work.}

\roles{All authors contributed to the conceptualization of the project and to the preparation of the manuscript. L.H.Y. and K.W.W. implemented the technical analyses of the model, detailed the physical proposal, and processed the numerical computations. J.B.N. contributed to the technical analyses and numerical computations. J.L. and A.F. designed the research framework, coordinated the program of activities, and edited the final version.}

%%%%%%%%%%%%%%%%%%%%%%%%%%%%%%%%%%%%%%%%%%%%%

%-----------------------------------%

\appendix

\section{\label{appA0} Hamiltonian derivation}
In this section, we provide details about the derivation of Hamiltonian \eqref{H1}. We start from the extended Bose-Hubbard model
\begin{eqnarray}
H = \frac{U_0}{2}\sum_{i=1}^4N_i(N_i-1)+\sum_{i,j=1}^4\frac{U_{ij}}{2}N_iN_j 
+\sum_{i=1}^4\nu_iN_i-J\sum_{i<j}^4(a_i^\dagger a_j + a_j^\dagger a_i).\nonumber
\end{eqnarray}
The on-site interaction energy $U_0$ results from the contact interaction, characterized by the scattering length, and DDI between particles at the same site, which is strongly influenced by the geometry of the potential well. 
Here, we assume a spherical potential so that the energy contribution resulting from the on-site DDI is cancelled out, such that the interaction energy $U_0$ can be controlled only by the scattering length via Feshbach resonance. On the other hand, the interaction energy between particles at site $i$ and $j$ depends on the orientation of dipoles, which can be controlled by a magnetic field, and is given by    
\begin{eqnarray}
U_{ij} = \widetilde{U}_{ij}(1-3\cos^2\theta),\nonumber
\end{eqnarray}
where $\theta$ is the angle between the polarization direction and the relative position vector between site $i$ and site $j$, and $\widetilde{U}_{ij}$ is the interaction energy when $\theta = \pi/2$. Here, we assume the dipoles are oriented along the $z$ direction. Due to the $z$-axis symmetry of the arrangement of wells, the inter-site interaction energies satisfy $U_{14}=U_{24}=U_{34}$, and $U_{12}=U_{23}=U_{13}$, and the Hamiltonian can be written as
\begin{align}
H&=\frac{U_0}{2}\sum_{i=1}^4N_i(N_i-1)+U_{14}(N_1+N_2+N_3)N_4+U_{12}(N_1N_2+N_2N_3+N_3N_1)\nonumber\\
&\qquad -J[a_4^\dagger (a_1+a_2+a_3)+ (a_1^\dagger+a_2^\dagger+a_3^\dagger)a_4].\nonumber
\end{align} 
Using the identity
\begin{eqnarray}
\frac{1}{2}\sum_{i=1}^4N_i(N_i-1)=\frac{N(N-1)}{2}-(N_1+N_2+N_3)N_4
%\nonumber\\&&
-(N_1N_2+N_2N_3+N_3N_1),\nonumber
\end{eqnarray}
the Hamiltonian reduces to (up to a global constant)
\begin{align}
H&=(U_{14}-U_0)(N_1+N_2+N_3)N_4+(U_{12}-U_0)(N_1N_2+N_2N_3+N_3N_1)\nonumber\\
&\qquad-J[a_4^\dagger (a_1+a_2+a_3)+ (a_1^\dagger+a_2^\dagger+a_3^\dagger)a_4].\nonumber
\end{align}
We focus on the case where sites 1, 2, and 3 form a non-interacting subsystem that leads to the integrability condition
$U_{12}=U_0$, 
so that the Hamiltonian can be written (up to a global constant $(U_{14}-U_0)N^2/4$) as 
\begin{eqnarray}
H=U(N_1+N_2+N_3-N_4)^2-J[a_4^\dagger (a_1+a_2+a_3)+ (a_1^\dagger+a_2^\dagger+a_3^\dagger)a_4].\nonumber
\end{eqnarray}
where we define
$U=(U_0-U_{14})/4$.
Note that 
$U_{14} = \widetilde{U}_{14}(1-3\cos^2\theta) = 0$,
since $\cos\theta = 1/\sqrt{3}$ for a cubic lattice. Thus, the interaction energy parameter is given by
$U=U_0/4$.

\section{\label{appD0} Bethe Ansatz and the effective Hamiltonian}
In this section, we obtain the state of the system using the Bethe ansatz method, following the approach of \cite{Bennett2024, Bennett2023}, which generalizes Bethe Ansatz techniques previously employed in studies of integrable atomtronic switching, NOON-state generation, and atom interferometry~\cite{Wittmann2018, GrunNOON2022, GrunPRL2022}. We consider the state
\begin{eqnarray}
|\Psi\rangle = \prod_{n=1}^{N-l-m}(u_nb_1^\dagger +a_4^\dagger)|0,l,m,0\},\nonumber
\end{eqnarray}
where 
\begin{eqnarray}
|0,l,m,0\}=|0\rangle\otimes \frac{(b_2^\dagger)^l(b_3^\dagger)^m}{\sqrt{l!m!}}|0,0\rangle \otimes |0\rangle.\nonumber
\end{eqnarray}

It can be shown that
\begin{equation}
H|\Psi\rangle = E|\Psi\rangle
+4U\sum_{n=1}^{N-l-m}u_n^2\left[u_n^{-1}\left(N-1\right)-\frac{\sqrt{3}J}{4U}(u_n^{-2}-1)-\sum_{r\neq n}^{N-l-m}\frac{2}{u_n-u_r}\right]b_1^\dagger |\Psi_n\rangle\nonumber
\end{equation}
where
\begin{align}
E&= U(N-2l-2m)^2+\zeta(l-m)-\sqrt{3}J\sum_{n=1}^{N-l-m}u_n,\nonumber\\
|\Psi_n\rangle&= \prod_{j\neq n}^{N-l-m}(u_j b_1^\dagger +a_4^\dagger)|0,l,m,0\}.\nonumber
\end{align}

Thus, the Bethe state is an eigenstate of the Hamiltonian if the roots satisfy the Bethe equation
\begin{eqnarray}
u_n^{-1}\left(N-1\right)-\frac{\sqrt{3}J}{4U}(u_n^{-2}-1)=\sum_{r\neq n}^{N-l-m}\frac{2}{u_n-u_r}.\nonumber
\end{eqnarray}

Next, we separate the set of roots into a set of small roots ($O(J/U)$, $n=p+1,\cdots, N-l-m$) and large ($O(U/J)$, $n=1,\cdots, p$) roots, such that
\begin{eqnarray}
\sum_{n=1}^{N-l-m}u_n = \underbrace{\sum_{n=1}^{p}u_n}_{\text{large}} +\underbrace{\sum_{n=p+1}^{N-l-m}u_n}_{\text{small}}.\nonumber
\end{eqnarray}

Then, for a general Fock initial state $|n_1,n_2,n_3,n_4\rangle$, $n_1+n_2+n_3+n_4=N$, and following the method of~\cite{Bennett2023}, the Bethe equation provides an expression for the quantum state of the system, given by

\begin{equation}
|\Psi(t)\rangle=\sum_{r=0}^{N-n_4}\sum_{s=0}^{N-n_4-r}\mathcal{A}_{r,s}^{n_2,n_3,n_4}(t)\,|N-n_4-r-s,r,s,n_4\rangle,\nonumber
\end{equation}  
where
\vspace{3.0 mm}
%\begin{widetext}
\begin{align}
\mathcal{A}_{r,s}^{n_2,n_3,n_4}(t)&=\sum_{l=0}^{N-n_4}\sum_{m=0}^{N-n_4-l}\frac{e^{-it E_{n_4,l,m}}}{\mathcal{N}_{n_4,l,m}^2}(B_{r,s}^{N-l-m-n_4,l,m})^* B_{n_2,n_3}^{N-l-m-n_4,l,m},\nonumber\\
B_{r,s}^{a,b,c} &= \sqrt{\frac{C_{a,b}^{a+b+c}}{C_{r,s}^{a+b+c}3^{a+b+c}}}A_{r,s}^{a,b,c}, \qquad C_{p,q}^n=\frac{n!}{p!q!(n-p-q)!},\nonumber\\
A_{r,s}^{a,b,c}&=\sum_{p=0}^{a}\sum_{q = 0}^{a-p}\sum_{i=0}^{b}\sum_{j = 0}^{b-i}C_{p,q}^{a}C_{i,j}^{b}C_{r-p-i,s-q-j}^{c}\,\nu^{p+2i-q-2j-r+s},\nonumber
\end{align}
and the normalization factor is given by
\begin{eqnarray}
\mathcal{N}_{n_4,l,m}^2 =1+\frac{J^2}{U^2}(A_{n_4,l,m}^2+A_{n_4+1,l,m}^2), \qquad 
A_{n_4,l,m}=\frac{\sqrt{3}}{4}\frac{\sqrt{n_4(N+1-l-m-n_4)}}{N+1-2n_4}.\nonumber
\end{eqnarray}
The corresponding energy is given by
\begin{eqnarray}
E_{n_4,l,m} =U(N-2n_4)^2+\zeta(l-m)-\frac{3J^2}{4U}\left[\frac{n_4(N-l-m-n_4+1)}{N+1-2n_4}-\frac{(N-l-m-n_4)(1+n_4)}{N-1-2n_4}\right].\nonumber
\end{eqnarray}
Note that for the initial Fock state $|N,0,0,0\rangle$, the energy expression reduces to 
\begin{eqnarray}
E_{0,l,m}=UN^2-(\xi-\zeta)l-(\zeta+\xi)m+\frac{3J^2}{4U}\frac{N}{N-1},\nonumber 
\end{eqnarray}
where $l$ and $m$ are eigenvalues of conserved operators $Q_2$ and $Q_3$, respectively. Then, ignoring the global constant $\displaystyle UN^2+\frac{3J^2N}{4U(N-1)}$, we conclude that the effective Hamiltonian is given by
\begin{eqnarray}
H_{\rm eff}=(\zeta-\xi) Q_2-( \xi+\zeta) Q_3.\nonumber
\end{eqnarray}

%-------------------------------------------%

\section{\label{appE0} Von Neumann entropy}
In this section, we provide more details about the calculation of the Von Neumann entropy. First, we observe that the state of system~\eqref{psi} can be written as
\begin{eqnarray}
|\Psi(t)\rangle= \sum_{l=0}^N\sum_{m=0}^{N-l}\mathcal{A}_{l,m}|N-l-m,l,m,0\rangle,\nonumber
\end{eqnarray}
\begin{eqnarray}
\mathcal{A}_{l,m}(t)=\frac{1}{3^N}\sqrt{\frac{N!}{l!m!(N-l-m)!}}\sigma_1^{N-l-m}\sigma_2^l\sigma_3^m.\nonumber
\end{eqnarray}
Tracing the density matrix over the modes of subsystems of wells 1,2, and 4, we obtain
\begin{eqnarray}
\rho = {\rm Tr}_{124}|\Psi(t)\rangle\langle\Psi(t)| = \sum_{s=0}^N\gamma_s|s\rangle\langle s|,\nonumber
\end{eqnarray}
where $\displaystyle \gamma_s = \sum_{k=0}^{N-s} |\mathcal{A}_{k,s}|^2$.
Then, the Von Neumann entropy is given by 
\begin{equation}
    S = -\sum_{s = 0}^N\gamma_s\log\gamma_s.\nonumber
\end{equation}

%------------------------------%

\section{\label{appF0} Error formula calculation}
To calculate the analytic expression of the imbalance population and evaluate the sensitivity against the rotation speed, we consider the inverse transformation of operators given by
\begin{align}
a_1&=\frac{1}{\sqrt{3}}(b_1+b_2+b_3),\nonumber\\
a_2&=\frac{1}{\sqrt{3}}(b_1+\nu^{-1} b_2+\nu b_3),\nonumber\\
a_3&=\frac{1}{\sqrt{3}}(b_1+\nu b_2+\nu^{-1} b_3),\nonumber
\end{align}
Using the inverse transformation, we obtain
\begin{align}
N_2 - N_3
&= \frac{i}{\sqrt{3}} \Big[
(b_2^\dagger b_1 - b_1^\dagger b_2)
+ (b_1^\dagger b_3 - b_3^\dagger b_1)
+ (b_3^\dagger b_2 - b_2^\dagger b_3)
\Big].\nonumber
\end{align}
Then, 
\begin{align}
\mathcal{O} &= e^{it H_{\rm eff}}(N_2-N_3)e^{-it H_{\rm eff}}\nonumber\\
&= \frac{i}{\sqrt{3}}\left[(e^{-it\alpha_-}b_2^\dagger b_1-e^{it\alpha_-}b_1^\dagger b_2)+(e^{it\alpha_+}b_1^\dagger b_3-e^{-it\alpha_+}b_3^\dagger b_1) +(e^{-i2t\zeta}b_3^\dagger b_2-e^{i2t\zeta}b_3^\dagger b_2)\right],\nonumber
\end{align}
where we define $\alpha_\pm = \xi\pm \zeta$. Using the inverse transformation, we obtain the relevant terms

\begin{align}
\mathcal{O} &= I(t) N_1+\frac{i}{3\sqrt{3}}(Aa_2^\dagger a_1-A^*a_1^\dagger a_2+B a_3^\dagger a_1-B^* a_1^\dagger a_3)+\cdots,\nonumber\\
\mathcal{O}^2 &= [I(t)]^2 N_1^2+\frac{1}{27}\left(|A|^2 +|B|^2\right)N_1+\cdots,\nonumber
\end{align}
where $(\cdots)$ denotes terms that do not contribute to the final result, and we define

\begin{align}
I(t)&= \frac{2}{3\sqrt{3}}\left[\sin(2t\zeta)-2\sin(\zeta t)\cos(\xi t)\right],\nonumber\\
A &=e^{-it\alpha_-}\nu^{-1}-e^{it\alpha_-} - e^{-it\alpha_+}\nu + e^{it\alpha_+}+ e^{-2it\zeta}\nu- e^{2it\zeta}\nu^{-1},\nonumber\\
B &=e^{-it\alpha_-}\nu-e^{it\alpha_-} - e^{-it\alpha_+}\nu^{-1}+ e^{it\alpha_+}+ e^{-2it\zeta}\nu^{-1} - e^{2it\zeta}\nu.\nonumber
\end{align}\\ 
For the initial Fock state $|\Psi_0\rangle =|N,0,0,0\rangle$, we obtain
\begin{align}
\langle N_2-N_3\rangle &= \langle \mathcal{O}\rangle = I(t)N\nonumber\\
\langle (N_2-N_3)^2\rangle &= \langle \mathcal{O}^2\rangle = [I(t)N]^2+G(t)N\nonumber
\end{align}
where
\begin{align}
G(t)&=\frac{1}{27}\left(|A|^2 +|B|^2\right)\nonumber\\
&=\frac{2}{27} \left\{\cos [2 t (\zeta -\xi )]-\cos [t (\zeta -\xi )]-2 \cos [t (3 \zeta -\xi )]-\cos [t (\zeta +\xi )]\nonumber\ +\cos [2 t (\zeta +\xi )] \right. \nonumber\\
&\qquad\qquad \left.-2 \cos [t (3 \zeta +\xi )] -\cos (2 \zeta  t)+\cos (4 \zeta  t)-2 \cos (2 \xi  t)+6\right\}.\nonumber
\end{align}
At $t = \tau = \pi/|\xi|$ we obtain,
\begin{align}
\langle N_2-N_3\rangle &= \frac{2N}{3\sqrt{3}}\left[\sin\left(\frac{2\pi\zeta}{\xi}\right)+2\sin\left(\frac{\zeta \pi}{\xi}\right)\right],\nonumber\\
G(\pi/\xi)&=\frac{8 }{27 }\left[\cos \left(\frac{\pi  \zeta }{\xi }\right)+2\right] \cos ^2\left(\frac{3
   \pi  \zeta }{2 \xi }\right).\nonumber
\end{align} 
Using the above results yields
\begin{align}
\Delta\langle N_2-N_3\rangle &=\sqrt{\langle (N_2-N_3)^2\rangle-(\langle N_2-N_3\rangle)^2} \nonumber\\
&= \sqrt{\frac{8 N}{27 }}\sqrt{2+\cos \left(\frac{\pi  \zeta }{\xi }\right)}\, \cos \left(\frac{3
   \pi  \zeta }{2 \xi }\right).\nonumber
\end{align}
Defining the dimensionless parameter directly related to angular velocity $\alpha = \zeta/J$,
the sensitivity of the rotation protocol can be evaluated using the error formula
\begin{align}
\Delta \alpha &= \frac{\Delta\langle N_2-N_3\rangle }{\left|\partial_{\alpha}\langle N_2-N_3\rangle\right|}\nonumber\\
&= \frac{\xi  }{2 \pi \sqrt{2N}J}\left[\sqrt{2+\cos \left(\frac{\pi  \zeta }{\xi }\right)} \,\sec \left(\frac{\pi 
   \zeta }{2 \xi }\right)\right],\label{deltaalpha}
\end{align}
where, for $0\leq \zeta\leq \xi/3$,
\begin{eqnarray}
\sqrt{3}\leq \sqrt{2+\cos \left(\frac{\pi  \zeta }{\xi }\right)} \,\sec \left(\frac{\pi 
   \zeta }{2 \xi }\right)\leq \sqrt{3+\frac{1}{3}}.\nonumber
\end{eqnarray}

\section{\label{appCR} Quantum Fisher information and the Cramér-Rao bound}   
 The quantum Fisher information for a pure state is given by \cite{Pezze2018}
\begin{eqnarray}
F_Q = 4(\langle \partial_\alpha \Psi|\partial_\alpha \Psi\rangle-|\langle\partial_\alpha \Psi|\Psi\rangle|^2),\nonumber
\end{eqnarray}
where the state of the system is given by
\begin{eqnarray}
|\Psi\rangle = \exp(-i\tau H)|N,0,0,0\rangle = \exp(i\tau \alpha J\mathcal{J})|\widetilde{\Psi}\rangle,\nonumber
\end{eqnarray}
where we define
\begin{eqnarray}
|\widetilde{\Psi}\rangle= \exp(-i\tau H|_{\zeta=0})|N,0,0,0\rangle.\nonumber
\end{eqnarray}
Derivation of the above expression provides
\begin{eqnarray}
|\partial_\alpha\Psi\rangle = \partial_\alpha |\Psi\rangle = i\tau  J\mathcal{J}|\Psi\rangle,\nonumber
\end{eqnarray}
from which we obtain
\begin{align}
\langle\partial_\alpha\Psi|\partial_\alpha\Psi\rangle &= \tau^2 J^2 \langle \Psi|\mathcal{J}^2|\Psi\rangle,\nonumber\\
\langle\Psi|\partial_\alpha\Psi\rangle &= i\tau J \langle \Psi|\mathcal{J}|\Psi\rangle,\nonumber
\end{align} 
such that
$
F_Q = 4\tau^2 J^2\,{\rm Var}(\mathcal{J})$,
where the variance reads
\begin{eqnarray}
{\rm Var}(\mathcal{J}) = \langle \Psi|\mathcal{J}^2|\Psi\rangle - |\langle \Psi|\mathcal{J}|\Psi\rangle|^2.\nonumber
\end{eqnarray}
Since $[H,\mathcal{J}] = 0$, the variance can be calculated from the initial state:
\begin{align}
{\rm Var}(\mathcal{J}) = \langle N,0,0,0|\mathcal{J}^2|N,0,0,0\rangle - |\langle N,0,0,0|\mathcal{J}|N,0,0,0\rangle|^2=\frac{2N}{3} \nonumber
\end{align}
Thus, the quantum Fisher information is 
$F_Q = 8\tau^2 J^2 N/3$
providing the Cramér-Rao bound \cite{Pezze2018}
\begin{eqnarray}
(\Delta \alpha)_{\rm CRB} = \frac{1}{\sqrt{F_Q}} = \frac{\sqrt{3}}{2\tau J\sqrt{2N}} = \frac{3\sqrt{3}J}{8\pi U\sqrt{2N}(N-1)}.\label{crb}
\end{eqnarray}

%-------------------------------%

\bibliographystyle{iopart-num}
\bibliography{biblio}

\end{document}